\documentclass[11pt,twocolumn]{article}

\ifx\pdfoutput\undefined
\usepackage{graphicx}
\else
\usepackage[pdftex]{graphicx}
\fi
\graphicspath{{figures/}}
\usepackage{url}       % Written by Donald Arseneau
\usepackage{amsmath}   % From the American Mathematical Society
\usepackage[margin=1in]{geometry}

\begin{document}

\date{}

% paper title
\title{A Distributed Economics-based Infrastructure for Utility Computing}

\author{Michael Treaster, Nadir Kiyanclar, Gregory A. Koenig, William Yurcik\\
National Center for Supercomputing Applications (NCSA)\\
University of Illinois\\
Email: \{treaster, nadir, koenig, yurcik\}@ncsa.uiuc.edu}

\maketitle

\begin{abstract}
Existing attempts at utility computing revolve around two approaches.  The first consists of proprietary solutions involving renting time on dedicated utility computing machines.  The second requires the use of heavy, monolithic applications that are difficult to deploy, maintain, and use.  

We propose a distributed, community-oriented approach to utility computing.  Our approach provides an infrastructure built on Web Services in which modular components are combined to create a seemingly simple, yet powerful system.  The community-oriented nature generates an economic environment which results in fair transactions between consumers and providers of computing cycles while simultaneously encouraging improvements in the infrastructure of the computational grid itself.
\end{abstract}

\section{Introduction}

Over the past decade, the declining price and increasing processing power of computing hardware has allowed cluster\footnote{We use the term ``cluster'' to refer to any kind of supercomputing resource.} systems to become much more common.  Although these systems are inexpensive compared with earlier incarnations of clusters, they still represent a considerable investment by an organization.  Many organizations have highly fluctuating demand for supercomputing resources and cannot justify the purchase of hardware to satisfy their peak demand.  Other organizations possess systems adequate to handle their peak demand, but these systems are then left idle when demand drops.  

Utility computing, also known as on-demand computing, is a model of computing that allows organizations or individuals to accommodate fluctuating demand for computing resources.   Users are able to acquire extra processing power as it is needed, obviating the need to purchase and maintain expensive hardware in order to meet maximum computing demands.

There are many commercial initiatives that attempt to provide utility computing services~\cite{hp-utility, ibm-utility, sun-utility}.  Some approaches establish a centralized service provider similar to traditional utilities like electricity or water.  In these approaches, a provider organization sets up supercomputing systems which are then leased to outside users as needed.  Another common approach is to deploy large systems to user sites, then unlock computing resources on these systems as they are needed by the user.  These solutions have the disadvantage of tying users to a particular vendor and a particular platform, making it more difficult for users to get a fair price on computation time.

Research institutions are developing projects related to specific aspects of utility computing~\cite{oasis04:nimrodg, cod, punch, condor}.  While commercial solutions are typically driven by providing users with compute cycles as they are needed and maximizing corporate profits, academic research tends to focus on using existing supercomputing systems as efficiently as possible.  This is typically accomplished by distributing jobs across many systems run by many organizations and allocating new jobs to the systems with the lightest loads.  

We present Superglue, a software infrastructure for utility computing based on~\cite{koenig04design}.   It employs a distributed network of Web Services to automatically bring together jobs submitted by users with clusters willing and able to execute them.  Each Web Service encapsulates a part of the functionality required to support the infrastructure as a whole, and it communicates with other Web Services to access functionality it does not provide for itself.  The infrastructure is cross-platform and sits on top of existing supercomputing hardware and software.  It does not have any requirements concerning the specifications of the system, such as requiring a particular batch scheduler or processor architecture.   

The Superglue infrastructure leverages the idle cycles on existing systems and allocates them to users as they are needed.  Unlike many solutions that attempt to allocate cycles to load systems equally, Superglue is designed to generate an economic environment where computing cycles are traded as a commodity, either through a bartering system for other computing cycles or in exchange for real money.  Not only can users purchase computing cycles as they are needed, but owners of clusters are also able to profit from idle processor time.  This economy allows market forces to regulate the price and availability of computing time and scheduling features within the Superglue system.

\section{Related Work}

\subsection{Globus}
The Globus Toolkit~\cite{oasis04:globus} is a set of utilities meant to aid in the setup and deployment of a Grid infrastructure and Grid enabled applications.  Globus is not intended to be a monolithic Grid solution, but rather consists of a set of interoperable tools on top of which Grid-aware applications can be built.  Globus is organized around three ``pillars'', each of which is represented by a member of the Globus Toolkit: the Grid Allocation Resource Manager (GRAM) for resource management, GridFTP for data transfer, and the Monitoring and Discovery System (MDS) for information management and resource discovery.  All of these components rest on a foundation of strong encryption, provided by the Grid Security Infrastructure (GSI).   

\subsection{Moab Grid Scheduler}
The Moab Grid Scheduler~\cite{oasis04:moab} (Moab) is a tool which is intended to facilitate the creation of a Grid.  Moab is a meta-scheduler which runs on top of lower level schedulers.  Individual organizations comprising a grid run the Maui cluster scheduler along with a resource manager like PBS~\cite{pbs}.  Maui enhances the resource manager by providing advanced quality of service and reservation capabilities, and allows communication and coordination with Moab.  Moab then provides a Grid-level global submission queue, and can use reservations to allocate resources across clusters if desired.    

Globus and Moab seek to solve the problem of coordinating distributed resource allocation among cooperating organizations which make up a Grid~\cite{oasis04:gridanatomy}.  These systems and Superglue have similar goals in that they all seek to increase the overall utilization efficiency of a network of supercomputing resources.  One feature which sets Superglue apart from these other systems is the leveraging of market forces to determine resource allocation.  As is described in Section~\ref{implementation}, this will require the integration of a banking system into Superglue.  Several other projects address the issue of market forces in Grid resource allocation.  GridBank~\cite{oasis04:gridbank} is a proposal for a Grid Accounting Services Architecture (GASA) to provide a Grid-wide accounting and banking infrastructure.  The Gold Accounting Manager~\cite{oasis04:gold} may, in the future, enable market driven resource brokering in the Moab Grid Scheduler.

\subsection{Faucets}
Faucets~\cite{faucets} is a project whose goal is to create a market economy for computing cycles.  Faucets has been developed in conjunction with the Charm~\cite{CharmppPPWCPP96, CharmppOOPSLA93} system, and much of the work on the project to date has focused on the development of an adaptive scheduler which can dynamically resize Charm-based batch jobs.  Faucets features an architecture similar to Superglue, whereby clients negotiate with a broker to determine which resource will run a given job.  It is our hope Superglue will benefit in terms of extensibility and interoperability by basing its architecture on composable Web Services.

\section{Economic Environment}
\label{economics}

Superglue will create a computational economy for compute cycles, treating processing time as a product that is produced by clusters during idle periods and sold to users when they need additional processing capabilities.  This idea is not new.  Many projects in the past have suggested trading computation time as a commodity and relying on market forces to regulate price~\cite{abramson02nimrodg, cpm01, faucets, nisan98popcorn, waldspurger92spawn, wolski00gcommerce}.  These designs, like Superglue, use supply and demand to ensure fair prices and an efficient market.

Clusters compete against each other in selling computation time to buyers.  They can differentiate their products based on hardware and software specifications, scheduling features, quality of service, and price.  Providing exceptional performance in one or more of these areas can provide a cluster with a competitive advantage over other sellers, allowing it to sell its computation time at a higher price while still attracting buyers.  

Superglue allows clusters to use custom bidding algorithms for generating the price they offer to buyers.  A cluster system can customize its price based on its current job load and on the features and quality of service required by a user.  This enables a system to pass on the cost of more expensive features to the users who employ them, and it also enables a system to affect its job load by considering its current load when generating a price.  

The Superglue infrastructure seeks to reduce the barriers to entry for potential sellers of computation time.  It allows anyone to easily install a lightweight front-end service on their cluster to automatically make the cluster's spare computation time available to potential buyers.  This ease of entry to the market should result in a sufficiently large population of suppliers for economic principles to take effect.

Supply and demand will encourage providers to become more efficient so they can sell computation time at a lower price or so they can upgrade the performance, features, and quality of service they are able to provide.  In this way, Superglue encourages improvement of the grid infrastructure.

\section{Superglue Vision}
\label{vision}
\begin{figure*}
\centering
\includegraphics[scale=.8]{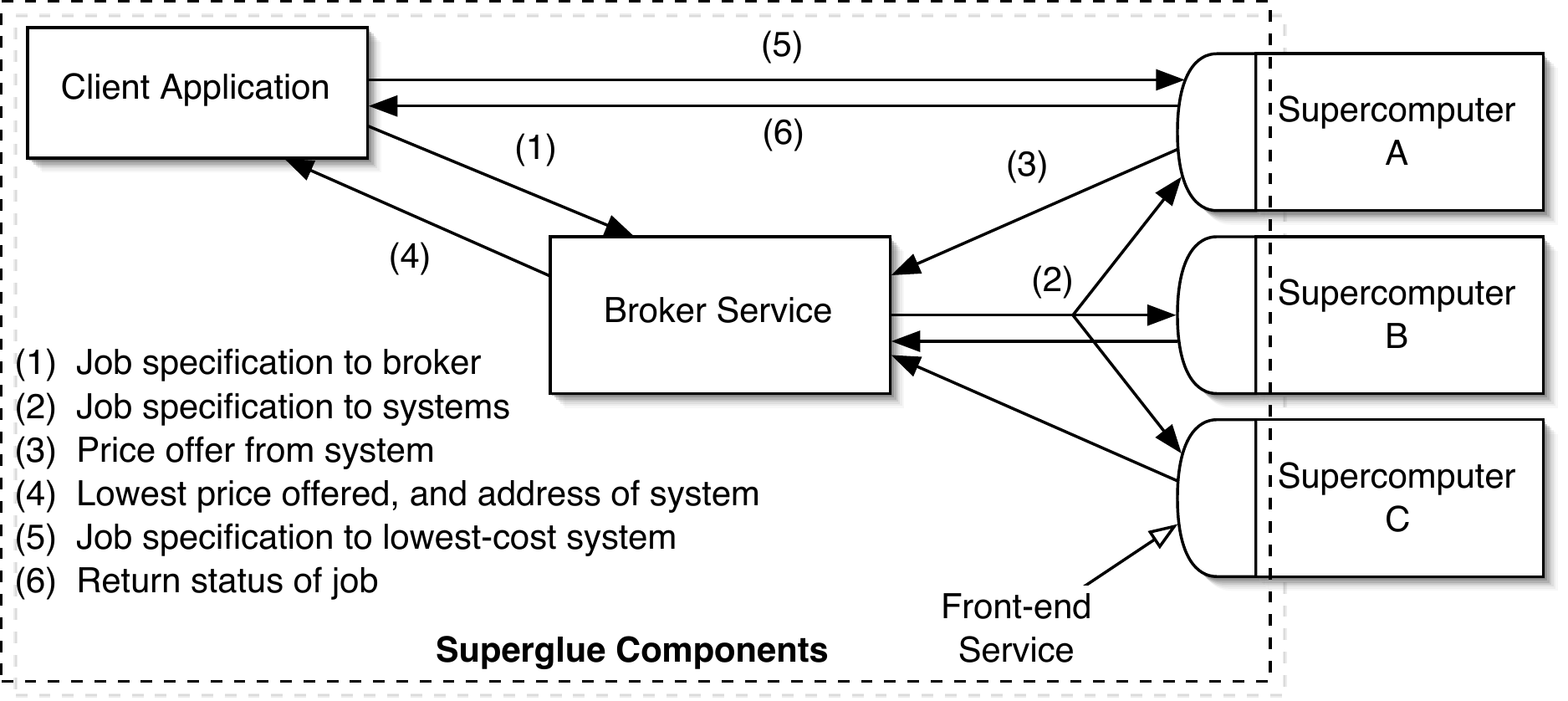}
\caption{Architecture Diagram for Superglue Prototype}
\label{arch-proto}
\end{figure*}

Superglue is a software infrastructure for utility computing.  It relies on existing supercomputing resources to provide the hardware aspects of the system rather than mandating the deployment of new supercomputing resources.   Superglue encourages both users and clusters to participate in the network by providing a valuable service to each party.  

The system matches each submitted job with the cluster willing to execute the job for the lowest cost to the user.  This allows users to obtain additional computing cycles during spikes in their workload by submitting their jobs to external systems with available computing time.  It simultaneously provides clusters with the opportunity to reduce their idle time while earning monetary compensation.  This provides a tangible benefit to supercomputing systems while also reflecting the goals of utility computing, where customers pay only for the computing time they use.

The underlying infrastructure of Superglue is intended to be nearly transparent from a user's perspective.  Like traditional job submission tools that reside on a cluster~\cite{ll, lsf, pbs}, it submits a job using a simple command line program.  The exact syntax will be different from the tools that the user already knows, but it will look familiar.   Behind this facade of simplicity hides a substantial amount of complexity, all automated by Superglue.  

A user constructs a job specification to precisely describe the job to candidate cluster systems.  This description encompasses all aspects of the job, including the resources, time allocation, and quality of service it will require, as well as user-related information such as the type of authentication that should be used.  When the job is submitted, Superglue searches for computing systems that can fulfill these requirements and queries each system for the price it will charge to execute the job.  The system offering the lowest price while fulfilling the requirements is selected by Superglue to run the job.  The money requested by the system is deducted from the user's account and held in escrow until the job is completed.  The user's files for the job are staged to the selected cluster's local file system and the job is submitted.  When the job execution completes, Superglue confirms fulfillment of the job specification, then transfers the money in escrow to the cluster's account. 

The system is also intended to be nearly transparent from a cluster's perspective.  The cluster is not required to run specific, unusual software, such as a certain scheduler.  The Superglue infrastructure sits on top of individual clusters, and jobs are submitted to the queues already in place on the system.  All that is required is that the cluster run an additional, light weight, front-end service that allows it to interact with the rest of the Superglue system.  This inobstrusiveness will make it easier for systems to join the Superglue network. 

Since Superglue relies on the scheduler that a supercomputing system chooses to use, it cannot guarantee that all computing resources in the network will be able to provide specific features such as deadlines or time reservations.  However, since each system can use a front-end service specific to the scheduler it is using, the front-end can surface such features to users of the Superglue network if the system administrator chooses to do so.  If a user requires a particular feature and a system does not support it, that system will not be selected to run the job.

\section{Implementation}
\label{implementation}

Superglue consists of a large collection of small Web Services, each providing a composable piece of functionality to the overall system.  Each service exposes an XML-RPC~\cite{xmlrpc} interface to share its functionality with the rest of the infrastructure.  The particular Web Services that are contacted by a component of the infrastructure are encoded in a configuration file for that component.  This allows a user to easily plug his application into a new implementation of an external component, should one become available.  Additionally, the Superglue infrastructure can be easily expanded by creating new clients that plug into existing Superglue services.  

Each Superglue component has been implemented in Python due to its capacity to facilitate rapid development of complex software with relatively little code, and for its excellent support for creating Internet and Web-based applications.  As Superglue becomes more mature, components can be rewritten in other languages as demanded by performance, security, or other concerns.   

The distributed nature of the infrastructure allows refinement to occur transparently without requiring users to reinstall every component of the system.  This characteristic, combined with the distributed nature, small size, and simplicity of individual components also makes it feasible for a third party to quickly develop substitute implementations of provided services.

\subsection{Core Components}

We have identified and implemented three core components of the Superglue infrastructure:  the client program, the broker service, and the cluster front-end service.  These pieces are the bare essentials required for a user to submit a job to an unspecified supercomputing resource.  Each component is simple by itself, but when composed together they form the basis for a powerful utility computing platform.  We describe these components in greater detail below, and we show their relationships in Figure~\ref{arch-proto}. 

We make two strong assumptions in this prototype implementation.  First, we assume the user has a login name and password that are the same across all candidate supercomputing systems.  Second, we assume that all necessary files related to the job already exist on the selected cluster in a known location.  These assumptions are not scalable to the widespread deployment of the system, and addressing these issues will be a high priority task as we extend the system further. 

\subsubsection{Client Application}
\label{subsec-client}
The client program is the executable employed by the user to interface with the system.  It is intended to make the use of Superglue as simple as possible by making it resemble the use of existing cluster batch submission tools.  

The user specifies the characteristics of a job either in a script file or at the command line.  The client program first contacts the broker service (Section~\ref{subsec-broker}) to find the lowest-cost cluster for the job that the user wants executed.  Then it directly contacts the front-end service running on that system (Section~\ref{subsec-frontend}) and sends it the job specification.  The front-end submits the job described by the specification to the job queue in the name of the user.

\subsubsection{Broker}
\label{subsec-broker}
A broker is an agent responsible for negotiating a contract between two third parties.  In the case of Superglue, the broker acts as a middleman between users and clusters, finding the cluster willing to run a user's job for the lowest cost.  It receives a job specification from the client program (Section~\ref{subsec-client}) and queries each cluster front-end service (Section~\ref{subsec-frontend}) for that cluster's price to execute the job.  The broker selects the cluster offering the lowest price and returns the address for that cluster to the client program. 

The Superglue design does not preclude the existence of multiple instances of the broker service.  A user or organization could easily run a private broker to reduce dependence on external components of the Superglue system.  

\subsubsection{Cluster Front-end Service}
\label{subsec-frontend}
Each participating cluster in the Superglue network runs a service to interface the cluster with the rest of the network.  This service has two purposes.  

First, it is responsible for generating a price when the broker (Section~\ref{subsec-broker}) asks the cost of executing a particular job specification.   The default implementation for the bidding algorithm returns a cost proportional to the cluster's current job load, such that a system with a lighter load will return a lower price.  This implementation will encourage users to choose systems with lighter loads, leading to an even distribution of job scheduling across the Superglue network and minimizing the idle time of any particular system.  However, a wide variety of other viable pricing algorithms exist, and supercomputing organizations can use whichever policy best maximizes their profits.

The second purpose of the cluster front-end service is to submit a job to the cluster on behalf of a user after it has been selected to run the job.  It is responsible for decoding the provided job specification and generating the correct information to submit the job to the cluster's local job queue using its own scheduling software.  Once the correct scripts and command sequences are generated, the front-end service submits the job.

\subsection{Future Extensions}
We have identified a variety of Web Services that are required to extend Superglue beyond the bare-bones framework that has been implemented to date.  We have focused our attention on identifying components that will be valuable for developing a robust, full-featured platform for utility computing.  It is likely there are other services or client programs that could exploit and supplement the Superglue infrastructure in ways we have not foreseen.

\begin{itemize}
\item \textit{Authentication Manager} -- Superglue could conceivably accept a variety of types of credentials.  The authentication manager will contact the appropriate credential authority to verify a user's login before accepting a job submission to the system.  It acts as a front-end to all credential authorities in order to abstract credential validation for the rest of the Superglue network.  This component might be heavily based on existing technologies such as MyProxy~\cite{myproxy} from the Globus project. 

\item \textit{Account Manager} -- This service will store account information for users and clusters.  Account information could consist of personal information, authentication information, and financial information.  It will also enable a prospective user to create an account with little or no assistance from an administrator, allowing new users to get up and running with minimal difficulty.

\item \textit{Bank} -- Superglue requires an automated mechanism for exchanging money to allow users to pay supercomputing systems for computation time.  The bank service is responsible for enabling financial transactions between users.  It might maintain the accounts itself, or it might act as a front end to an external Web Service, such as PayPal or a traditional bank.

\item \textit{Globus, Condor, et al Front-ends} -- This is a collection of services, with each individual service acting as a front-end to networks based on other grid computing or utility computing solutions such as Globus or Condor.  These services will allow existing grid networks to participate in the Superglue network as a supercomputing resource, capable of placing bids and executing jobs in exchange for money.  This will allowing Superglue to leverage existing grid infrastructure.  

\end{itemize}

The elegance of the Web Service approach is that each piece of functionality can be implemented individually.  The system can be built up piece by piece, gaining features and robustness as it grows, but providing functionality to users before it is complete.  We expect that additional features in the form of new Web Services and client applications will continue to arise indefinitely.

\section{Conclusions}
\label{conclusions}

Utility computing is an important, unrealized step in providing supercomputing capabilities to users with projects of any scale.  We believe that a distributed, economics-based system is the best way to deploy a robust, flexible infrastructure to fulfill this goal.  It is lightweight, easy to deploy, and easy to use.  This makes it simple for anyone to participate in the system, either as a buyer or as a seller.  

The distributed, composable nature of the infrastructure has a variety of advantages.  It allows for rapid development of the system, since the architecture is logically separated into simple, independent components.  Individual components can be replaced with improved versions without requiring software upgrades by every participant in the system.  Also, third parties can offer their own versions of individual components either for private use or by the community at large, allowing for community-driven improvement of or extensions to the Superglue infrastructure.

Clearly there are many technical issues that remain to be solved.  Not all questions have been answered nor solutions devised.  How are job files efficiently staged to the selected cluster and how is output data retrieved?  What kind of security is necessary to protect supercomputing systems and Superglue components?  What forms of authentication are necessary, and how should they be implemented?  With the modular Superglue architecture, each of these questions can be addressed by a particular Web Service dedicated to solving each specific problem.  As these solutions are implemented, they can be integrated into the larger Superglue network with minimal difficulty.  

We expect that the introduction of competition and supply and demand to utility computing will prove beneficial to the community.  Economic pressures will regulate the availability of features as well as the price at which features are available.  This will encourage improvements to the grid infrastructure as clusters improve both their hardware and software in order to remain competitive in the Superglue environment.

\end{document}